\documentclass[sigconf]{acmart}

\usepackage{booktabs} 
\usepackage{caption}
\usepackage{subcaption}
\usepackage[utf8]{inputenc}
\usepackage{listings}
\usepackage{mathtools,xparse}
\usepackage{float}
\setcopyright{rightsretained}
\usepackage{balance}

\balance

\acmDOI{10.475/123_4}

\acmISBN{123-4567-24-567/08/06}
\acmConference[DS+J'17]{DATA SCIENCE + JOURNALISM @ KDD 2017}{August 2017}{Halifax, Canada} 
\acmYear{2017}
\copyrightyear{2017}
\acmPrice{15.00}

\newcommand{\wagner}[1]{{}}

\newcommand{\manoel}[1]{{}}

\begin{document}

\title{``Everything I Disagree With is \#FakeNews'':
\newline
Correlating Political Polarization and Spread of Misinformation}

\author{Manoel Horta Ribeiro, Pedro H. Calais, Virgílio A. F. Almeida, Wagner Meira Jr.}
\orcid{1234-5678-9012}
\affiliation{%
  \institution{Universidade Federal de Minas Gerais}
  \city{Belo Horizonte} 
  \state{Minas Gerais} 
  \country{Brazil}
}
\email{{manoelribeiro, pcalais, virgilio, meira}@dcc.ufmg.br}
\renewcommand{\shortauthors}{Manoel Horta Ribeiro. et al.}

\begin{abstract}
An important challenge in the process of tracking and detecting the dissemination of misinformation  is to understand  the gap in the political views between people that engage with the so called "fake news". 
A possible factor responsible for this gap is opinion polarization, which may prompt the general public to classify content that they  disagree or want to discredit as fake. In this work, we study the relationship between political polarization and content reported by Twitter users as related to "fake news”. We investigate how polarization may create distinct narratives on what misinformation actually is. We perform our study based on two datasets collected from Twitter. The first dataset contains tweets about US politics in general, from which we compute the political leaning of each user towards the Republican and Democratic Party. 
In the second dataset, we collect tweets and URLs that co-occurred with "fake news" related keywords and hashtags, such as \#FakeNews and \#AlternativeFact, as well as reactions towards such tweets and URLs. 
We then analyze the relationship between polarization and what is perceived as misinformation, and whether users are designating information that they disagree as fake.
Our results show an increase in the  polarization of users and URLs (in
terms of their associated political viewpoints) for information labeled with fake-news keywords and hashtags, when compared to information not labeled as "fake news".  We discuss the impact of our findings on the challenges of tracking "fake news" in the ongoing battle against misinformation.
\end{abstract}

%
%



\maketitle

\section{Introduction}

Online social networks have changed the news consumption habits of many, since they present news in a structure which differs dramatically from previous media technologies~\cite{marchi2012facebook}.
Online content can be spread with little or no filtering, and sources with negligible or unknown reputation may reach as many readers as established media outlets~\cite{allcott2017social}.
The profits derive mainly from clicks that attract the reader to the media's website, which increases the ``tabloidization" of the headlines~\cite{chakraborty2016stop}.
The information to which users are exposed is selected through  recommendation algorithms~\cite{liao2013beyond}, which may create  \textit{"filter bubbles"}, separating  users from information (and news) that disagrees with their viewpoints~\cite{pariser2011filter}.

In this context, two phenomena have been increasingly receiving attention due their potential impact on important societal processes~\cite{allcott2017social,bakshy2015exposure}: the rapid spread of a growing number of unsubstantiated or false information online~\cite{del2016spreading}, recently named as \textit{"fake news"}, and the increase of opinion polarization~\cite{allcott2017social,BalancingOpposingViews}. 
Previous studies suggest a dual interaction between the two. The polarized \textit{"echo chamber"} communities are more susceptible to the dissemination of misinformation~\cite{del2016spreading}. Conversely, misinformation plays a key role in creating polarized groups~\cite{zollo2015emotional}.
Another way these phenomena may interact is when users incorrectly classify news
as sources of misinformation simply due to disagreement, not because it reports actual false or imprecise facts~\cite{lewandowsky2012misinformation}. This behavior creates alternative narratives of what is actually fake, which depends on one's political ideology, and that, ultimately, make the line between biased and fake information more blurry. 

\begin{figure}[htp!]
    \centering
 \includegraphics[width=\linewidth]{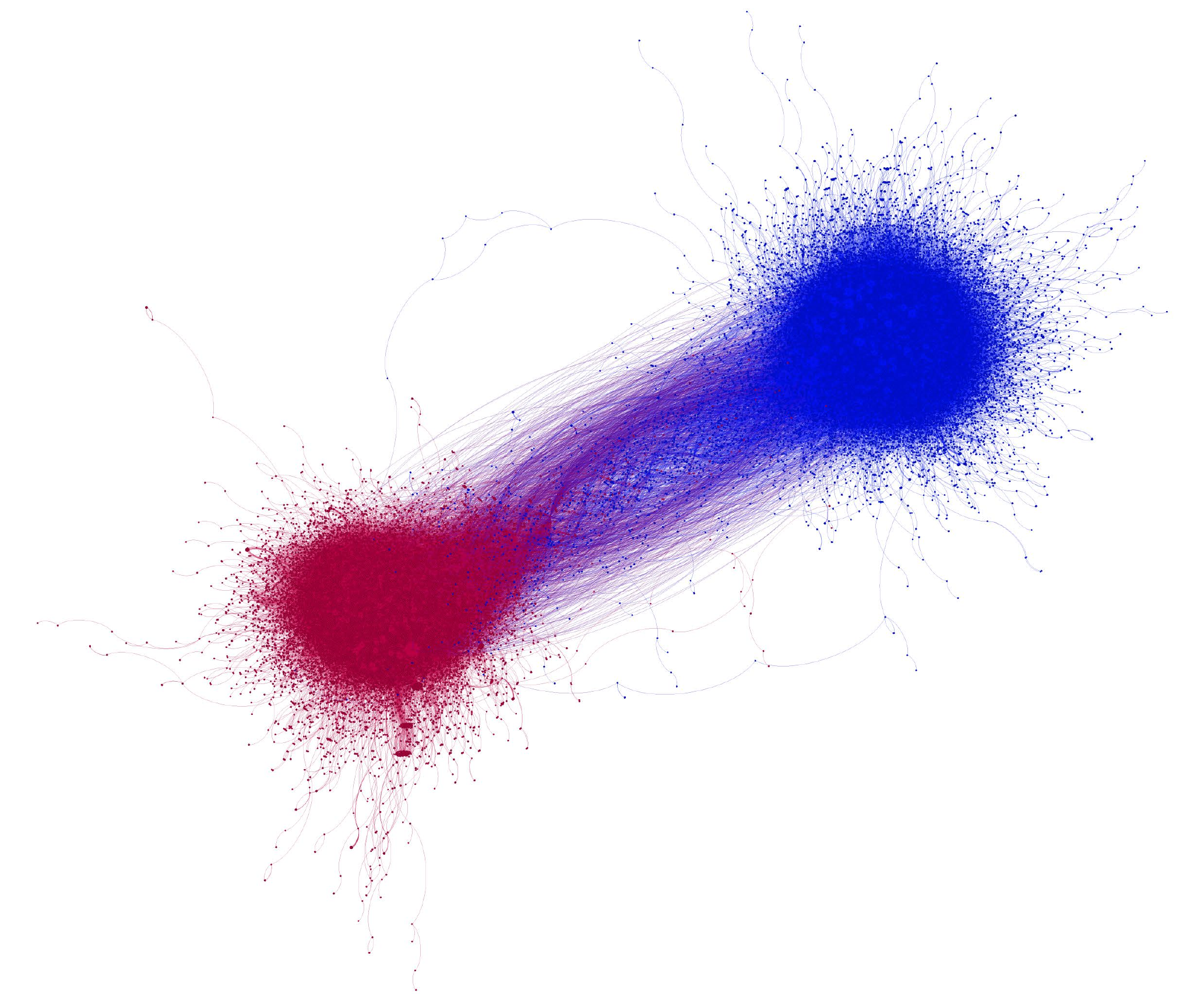}
  \caption{Network of retweets showing democrats (in blue) and republicans (in red) divided into two distinct communities. What is the impact of such polarization in what is perceived as "fake news"?}
       \label{fig:graph}
\end{figure}

In this paper we conduct an initial analysis on the relationships and interactions between polarized debate and the spread of misinformation on datasets collected from \textit{Twitter}\footnote{https://twitter.com/}. We examine the following research questions:

\begin{quote}
\textit{Q1: How is polarization quantitatively related to information perceived as or related to fake news?}
\end{quote}
\begin{quote}
\textit{Q2: Are users designating content that they disagree with as misinformation?}
\end{quote}

\noindent
We analyze a dataset composed of tweets on content associated with "fake news" and general tweets about U.S. politics.
Our methodology employs a community detection method designed to estimate the degree of polarization of each user~\footnote{We employ the term polarization for both the collective phenomena of opposition of opinions and to designate how much an individual user or URL leans towards a set of views or an ideology.} leaning towards
the Democratic or Republican parties, as depicted in Figure~\ref{fig:graph}. Based on these estimates, we correlate user polarization levels to their interactions with
\#FakeNews-related tweets and external URLs. We analyze how the polarization of such tweets and URLs is related to their popularity and to the frequency they are associated with the theme of misinformation by users. We also analyze the polarization differences between users who merely discuss politics versus those who engage with tweets and URLs related to fake news.

\begin{figure*}[!ht]
 \centering
 \includegraphics[width=\linewidth]{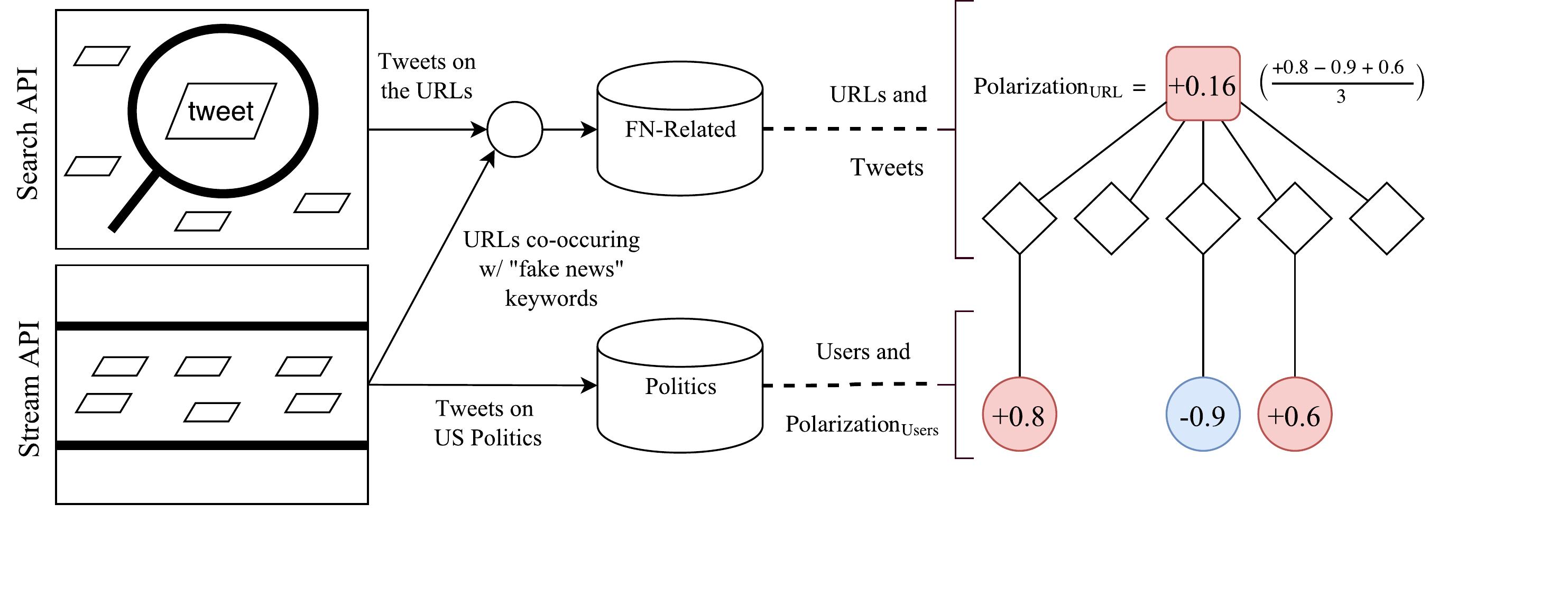}
 \caption{Methodology to collect the URLs flagged as fake news, general tweets that tweeted this URL, and general tweets on politics, and then build a dataset that encompasses the polarized reactions of users to an URL. We also exemplify how the calculation of the polarization of the URL is performed on the right-hand side, as it is further discussed in Section~\ref{sec:pol_url}}
 \label{fig:dataset_methodology}
\end{figure*}

Our data analysis process shows three main findings:

\begin{enumerate}
\item There is an increase in polarization on the URLs and users associated with fake-news related keywords and hash-tags; 
\item Polarized groups cite sources on their side of political spectrum to tag or condemn news and statements given by the other opposite group as fake;
\item Polarized users employ terms such as "fake news" to refer to content that they particularly disagree with.
\end{enumerate}

\noindent
We discuss the impact of these findings in the ongoing battle against the spread of online misinformation online. We suggest, for example, that approaches based on crowd-source~\cite{ratkiewicz2011detecting} to detect fake news may become biased towards political ideologies, once the narratives on what is fake seem to be quite different across groups with different ideologies.

The remainder of this paper is organized as follows. 
Section~\ref{related} reviews previous work on the spread of online misinformation and on opinion polarization.
Section~\ref{methodology} describes the methodology behind the data collection and data analysis processes.
Section~\ref{results} presents and discusses the results of our analysis.
Finally, Section~\ref{conclusion} concludes the paper and outline future research directions.

\section{Related Work}
\label{related}

The spread of online misinformation\footnote{Some sources distinguish misinformation and disinformation based on intention, we use misinformation for both.} and "fake news" has become an increasingly important topic for its possible impact in societal processes such as political elections
and public policies~\cite{SocialMediaAndFakeNews}. Closely tied to "fake news" are the so-called
"alternative narratives" such as conspiracy theories~\cite{starbird2017icwsm}. There is an ongoing effort in the research community to understand the spread and propagation of such kinds of content. Some of the approaches taken are to propose network diffusion models~\cite{tambuscio2015fact,wang20142si2r}, and analyzing network structural features of the propagation of misinformation in online social networks~\cite{lerman2010information, kwon2013prominent}.

Some effort has also been done to detect misinformation, including strategies that apply text-based methods~\cite{AutomaticDeceptionDetection}, fact-checking through knowledge graphs~\cite{ComputationalFactChecking}, crowd-sourcing solutions~\cite{ratkiewicz2011detecting}, and even verifying the authenticity of images spread online~\cite{pasquini2015towards}. There are also recent attempts on detecting the spread of misinformation using regularities on their propagation patterns through social networks, with limited success so far~\cite{aprilFool2017arxiv}. Besides that, previous studies also attempt to contain the spread of misinformation, finding near-optimal ways of disseminating information that may revert the damage caused by a rumor~\cite{budak2011limiting} or even strategies to clarify misinformation from a physicological perspective~\cite{lewandowsky2012misinformation}.

Another line of research surrounds the existence of political bots that spread misinformation. There are several studies on the impact of such bots in specific countries~\cite{forelle2015political, howard2016bots}, as well as more general studies on their strategies and particularities~\cite{woolley2016automating} and on methods for detecting them~\cite{ferrara2014rise}. The existence of such bots may have a strategic role in political debate, influencing, for instance, the trending hashtags on \textit{Twitter}~\cite{howard2016bots} and are a threat to healthy and productive political discourse. This makes the understanding and combat of such bots network an important challenge for the scientific community~\cite{subrahmanian2016darpa}.

The association between opinion polarization and "fake news" accusations has been suggested in the media
as strong; people would just label as "fake" any information or sources they do not support~\cite{slate,telegraph}.
From the sociological perspective, polarization may be formally understood as a state that ``refers to the extent to which
opinions on an issue are opposed in relation to some theoretical maximum'', and, as a process,
it is the increase in such opposition over time, causing a social group to divide itself into two sub-groups
with conflicting and antagonistic viewpoints regarding a topic~\cite{Law_Group_Polarization,Group_Polarization,US_More_Polarized}. Understanding polarization in online discussions and the social structures induced by polarized debate is important because polarization of opinions induces segregation in the society, causing people with different viewpoints to become isolated in islands where everyone thinks like them. Such "filter bubbles" caused  by social media
systems limits the exposure of users to ideologically diverse content, and is a growing concern~\cite{DoesFacebookIntroduceIdeologicalBias,BalancingOpposingViews}.
Recommendation algorithms in social media contexts may increase the scale of polarization even more, as they can automatically separate users from alternate viewpoints on polarized issues by not showing those on their feeds~\cite{pariser2011filter}.

This work is a first attempt to test the hypothesis that "fake news" narratives are correlated to political polarization. It differs from much of the preexisting work as it considers the possibility of individuals tagging as misinformation content that they disagree with. If significant, this adds another layer of complexity to the problem, as we need to distinct perceived "fake news" from what is actually misinformation.

\section{Methodology}
\label{methodology}

In this section we describe the methodology used to collect the data and the more important methods used in the data analysis process.

\subsection{Data Collection}

Our data collection strategy is shown in Figure~\ref{fig:dataset_methodology}.
We study two datasets in conjunction, both obtained from \textit{Twitter}. 
The first dataset was built to monitor narratives and discussions surrounding fake news. To do so, we performed
two simultaneous data collection efforts, using the Stream\footnote{https://dev.twitter.com/streaming/overview} and the Search APIs\footnote{https://dev.twitter.com/rest/public/search}. The Stream API allows you to gather large amounts of data currently being tweeted whereas the Search API allows you to search for tweets mentioning specific keywords (among countless other parameters). 

\begin{enumerate}
\item In the first step, we collect the stream of tweets containing the following keywords and hashtags from Twitter's Stream API:  

\begin{quote}
\noindent
\texttt{\{fakenews, 
\#\lowercase{fakenews}, 
fake-news, 
\#\lowercase{fake-news}, \\
posttruth,
\#\lowercase{posttruth},
post-truth, \\
\#\lowercase{post-truth},
alternativefact, \\
\#\lowercase{alternativefact}, 
alternative-fact, \\
\#\lowercase{alternative-fact}\}}
\end{quote}

\noindent
We then proceed to store the URLs being mentioned, whether it is an external URL or an URL to another tweet. For example:

\begin{quote}
\texttt{\textbf{External URL}: Trump Schools CNN Reporter in 1990 - Then Drops the Mic - Literally \{URL\} \#fakenews}
\end{quote}
\begin{quote}
\texttt{\textbf{Another Tweet}:  RT @\{User\}: This is an abuse of his office. \{Tweet\}}
\end{quote}

\item In the second step, the stored URLs are bufferized and consumed by another data collection process. Every $15$ minutes we use Twitter's Search API to extract general tweets that include the most relevant URLs stored, and meta-data about the users who tweeted about them. This let us capture the context surrounding the URLs in a broader scenario, with no necessary association to misinformation-related keywords or hashtags. We exemplify this with two tweets mentioning the same URL with different contexts:

\begin{quote}
\texttt{\textbf{Fake-news context}: Huffing ComPost is a joke. Nobody believes their \#fakepolls or \#fakenews. \#MAGA \{URL\}}
\end{quote}
\begin{quote}
\texttt{\textbf{Indirect Fake-news context}: Canadian views of U.S. hit an all-time low, poll shows, \{URL\}}
\end{quote}

\end{enumerate}

The goal of this double-step collection process is to build a more complete view of the fake
news debate on Twitter: we can see both users who are referring to a content (i.e. an URL or another tweet) as a potential
source of fake news, and users who are citing, propagating or interacting with the same content \emph{without} attaching to it the fake-news label. 

The second dataset we used was obtained by collecting tweets about US Politics in general from Twitter Stream API. We use keywords and hashtags 
such as \texttt{\{Hillary Clinton, \#\lowercase{POTUS}, Donald Trump, White House, Democrats, Republicans$\ldots$\}}.
The usefulness of this dataset on this work is to offer enough data to accurately compute the degree of polarization of users from the FN-dataset with respect
to their leanings towards Republicans and Democrats. This is explained in details in Section~\ref{sec:pol_usr}.

Some remarks on the methodology are that:

\begin{enumerate}
\item  Retweets and quote tweets are considered to be URLs to another tweets;
\item   The choice of $15$ minutes as a buffer time was imposed by limitations of Twitter's Search API;
\item  The data collection steps $1$ and $2$ were done from \textit{May 07 2017} to \textit{May 25 2017}, whereas the collection of step $3$ was done from \textit{August 2016} to \textit{May 2017}.
\end{enumerate}

Using these data sources, we are able to analyze  URLs that co-occurred with \#fakenews tags (obtained in step $1$), the associated reactions to this URL in the form of tweets (obtained in step $2$), and the polarization of some of the users who tweeted it (obtained in step $3$). This is depicted on the right-hand side of  Figure~\ref{fig:dataset_methodology}.

When we capture any URL contained in a tweet, we are capturing many different kinds of non-trivial interactions. For example, it has been shown that retweets may express disagreement~\cite{guerra2017antagonism}. These subtleties have little impact on our analysis, as we are interested not in the type of reaction that a user has, but whether users from differently polarized groups react at all to the same URLs.

\begin{table*}[ht]

\begin{tabular}{l|cccc|cc|cc}
\hline
           & \multicolumn{4}{c|}{General Statistics}                                & \multicolumn{2}{c|}{Shared Users} & \multicolumn{2}{c}{Shared Active Users} \\ \hline
Source     & \#users   & \multicolumn{1}{l}{\#active users} & \#tweets    & \#urls  & FN-Related       & Politics       & FN-Related         & FN-Related         \\ \cline{1-1}
FN-Related & 374,191   & 101,031                            & 833,962     & 109,397 & -                & $29.22\%$      & -                  & $37,61\%$          \\
Politics   & 4,164,604 & 247,435                            & 246,103,385 & -       & $2.62\%$         & -              & $15.72\%$          & -                  \\ \hline
\end{tabular}
\caption{General characterization of the data sources. The intersection between the \texttt{Politics} dataset and \texttt{FN-Related} is important as we use it to characterize the polarization of the users, and consequently of the URLs in the \texttt{FN-Related} datasets.}
\label{table:general}
\end{table*}

\subsection{Estimating User's Political Polarization}
\label{sec:pol_usr}

The main unit of information we want to correlate with fake news-related tweets is the degree
of polarization of Twitter users to each main side in US Politics -- Republicans and Democrats. Notice that as stated previously, we overload the word \textit{polarization} to denote how much an individual leans towards a set of views or an ideology.
There is a plethora of methods designed to classify the political leaning of social media users, which typically group themselves in well-separated communities~\cite{PoliticalPolarizationTwitter,QuantifyingPoliticalLeaningFromTweetsRetweets}. Although our methodology does not depend on the specific graph clustering algorithm,
finding communities on polarized topics is eased by the fact that it is usually
simple to find seeds -- users that are previously
known to belong to a specific community. In the case of the Twitter datasets we take into consideration, the official
profiles of politicians and political parties
are natural seeds that can be fed to a semi-supervised clustering
algorithm that expands the seeds to the communities formed around them~\cite{pcalaisKDD11,Snowden,KleinbergSmallSeedSets}.

We assume that the number of communities $K$ formed around a topic $T$ is known in advance and it
is a parameter of our method. To estimate user leanings toward each of the $K$ groups (K=$2$ for {Democrats, Republicans}),  we employ a label propagation-like strategy based on random walk with restarts~\cite{RWR}: a random walker
departs from each seed and travels in the user-message retweet bipartite graph by randomly choosing an edge to decide which node it should go next.
With a probability (1 - $\alpha) = 0.85$, the random walker restarts the random walking process from its original seed. As a consequence,
the random walker tends to spend more time inside the cluster its seed belongs to~\cite{pcalaisKDD11}.
Each node is then assigned to its closest seed (i.e., community), as shown
in the node colors in the sample of the graph displayed in Figure~\ref{fig:graph}.

The relative proximity of each node to the two sets of seeds yield a probability that this node belongs to 
each of two communities, and can be interpreted as an estimate of his or her political leaning. For instance, if proximity of node X to republican seeds is 0.01 and its proximity to democrat seeds is 0.04, the random-walk based community detection algorithm outputs that that node
belongs to the democrat community with 80\% of probability. Note that this modeling nature captures that
some nodes may be more neutral than others. For more details on the random walk-based community detection algorithm,
please refer to~\cite{pcalaisKDD11}. 

In our specific case study there are only two communities, thus we can define the polarization of a user $u$ with an assigned polarization value $v_u \in [0.5, 1.0]\cup [-1.0,-0.5]$ as a random variable $P_{u}: [-1, 1] \mapsto [0,1]$ such that:

\begin{equation}
P_{u} = 
\begin{cases}
2(-v_u + 0.5) \text{ if } u \in D \\
2(v_u - 0.5) \text{ if } u \in R
\end{cases}
\end{equation}

\noindent
Where $R$ and $D$ are the polarized groups of republicans and democrats. Notice that we are simply changing the domain of the value assigned by the polarization algorithm to a more intuitive one ($[-1, 1]$). We can further define the absolute user polarization as a random variable $A_{u}: [0, 1] \mapsto [0,1]$ such that:

\begin{equation}
A_{u} = |P_{u}|
\end{equation}

\subsection{Estimating URL's Political Polarization}
\label{sec:pol_url}

Another aspect of the data that needs to be modeled is the polarization of an URL, or in other words, how it reverberates through users in opposed polarized communities. We define the degree of polarization of an URL based on the polarization of users that reacted to it. For the sake of simplicity,
we consider as URLs any links external to Twitter or links to other tweets, and reactions are normal tweets, quotes, replies and retweets that interact with the URL.

We define the polarization of a URL $k$, given two polarized groups of users $R$ and $D$, as a random variable $P_{k} : [0, 1]^n \mapsto [0,1]$ that is the average of the polarization of users $P_u$ that reacted to it:

\begin{equation}
P_{k} = 
\frac{1}{n} 
\sum_{u \in \mathbf{U}(k)}^n 
P_u
\end{equation}

This calculation is depicted in the right-hand side of Figure~\ref{fig:dataset_methodology}. In a similar fashion as we did for the polarization of users, we can further define the absolute URL polarization as a random variable $A_{k}: [0, 1]^n \mapsto [0,1]$:

\begin{equation}
A_{k} = |P_{k}|
\end{equation}

\subsection{Domains and Impactful URLs}
\label{sec:word_clous_mpac}
An important part of our analysis is trying to find evidence that users are employing "fake news" related terms to express disagreement,
rather than a more factual lack of vericity in content they tweet about. To do so, we analyze the URL domains mentioned by each polarized sides in tweets associated to misinformation. We also qualitatively analyze the content of some of the URLs that generated the most significant reactions.

To generate the domains we parse the external URLs mentioned in the tweets. We then calculate the political polarization of each domain 
exactly like we do for full URLs. The wordclouds are generated for all the external URLs with absolute polarization $A_k$ bigger than $0.5$, one for each respective polarized group.
For the analysis of the content of the top  reacted URLs, we randomly select $75$ of the top $150$ URLs (tweets and external). Of those, we have equally sized stratum where $A_k$ belongs to intervals  $[0,0.32]$, $[0.33,0.66]$ or $[0.67,1]$. We then analyze the content of URLs for insights on different ways the fake news thematic may emerge.

\begin{figure*}
    \centering

 \includegraphics[width=\linewidth]{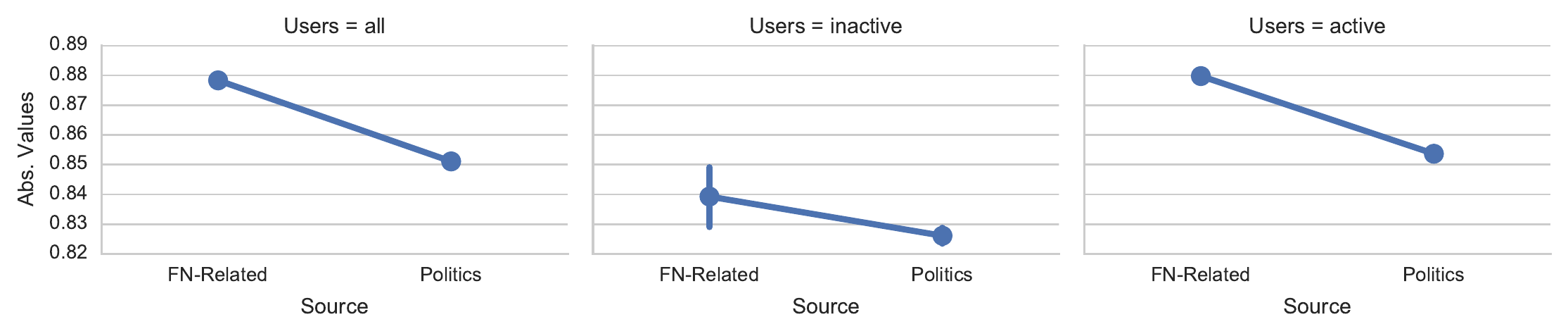}
     \caption{Average absolute user polarization for the users in the \texttt{FN-Related} and the \texttt{Politics} dataset. The error bars are the $95\%$ confidence intervals calculated using bootstrap. The increase in the polarization in the \texttt{FN-Related} dataset suggests that the theme of misinformation increases polarization in an already polarized topic (politics).}
    \label{fig:users_pol}
\end{figure*}

\section{Results}
\label{results}

We begin by characterizing the two datasets in terms of tweets, URLs and users, as depicted in Table~\ref{table:general}. Remember that the analysis concerning URLs are all performed using the tweets of the dataset we call \texttt{FN-Related} and the polarization of users of the dataset named \texttt{Politics}. 

The dataset sizes differ significantly, but the intersection amongst them grants us a significant number of the users to perform the analysis with ($29.22\%$ of the $374,191$ users in the \texttt{FN-Related} dataset). If we define the active users in the \texttt{Politics} as the smallest set of users responsible for 80\% of the tweets collected, we have that the intersection with the users from the \texttt{FN-Related} dataset grows significantly, increasing to $15.72\%$ from the original $2.62\%$.

\subsection{Polarization and Tweets}

We begin by analyzing the difference in polarization of the users in the \texttt{Politics} dataset and the \texttt{FN-Related} dataset. Notice that all the users that we know the polarization of in the second are also in the first. Figure~\ref{fig:users_pol} shows the average polarization in such datasets looking at all users, but considering active or inactive users.  The significant increase in polarization on the users that were associated with URLs that co-occurred with "fake news" related terms is an indication that the theme of "fake news" increases polarization in the already polarized discussion of politics.

\begin{figure*}
    \centering
          \begin{minipage}{.47\textwidth}
      \centering
      \includegraphics[width=\linewidth]{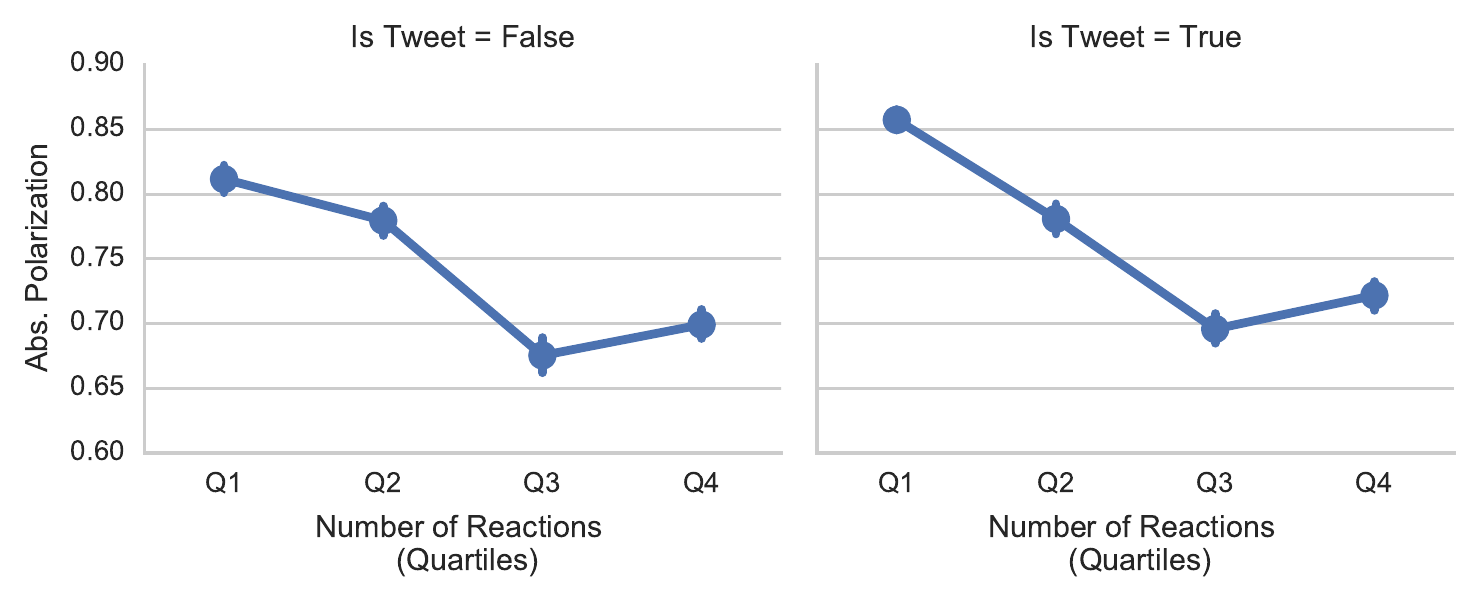}
      \caption{Average polarization per number of reactions to an URL (quartiles). Error bars represent $95\%$ confidence interval.}    
      \label{fig:rc_pp}

  \end{minipage}
  \hfill
    \begin{minipage}{.47\textwidth}
    \centering \includegraphics[width=\linewidth]{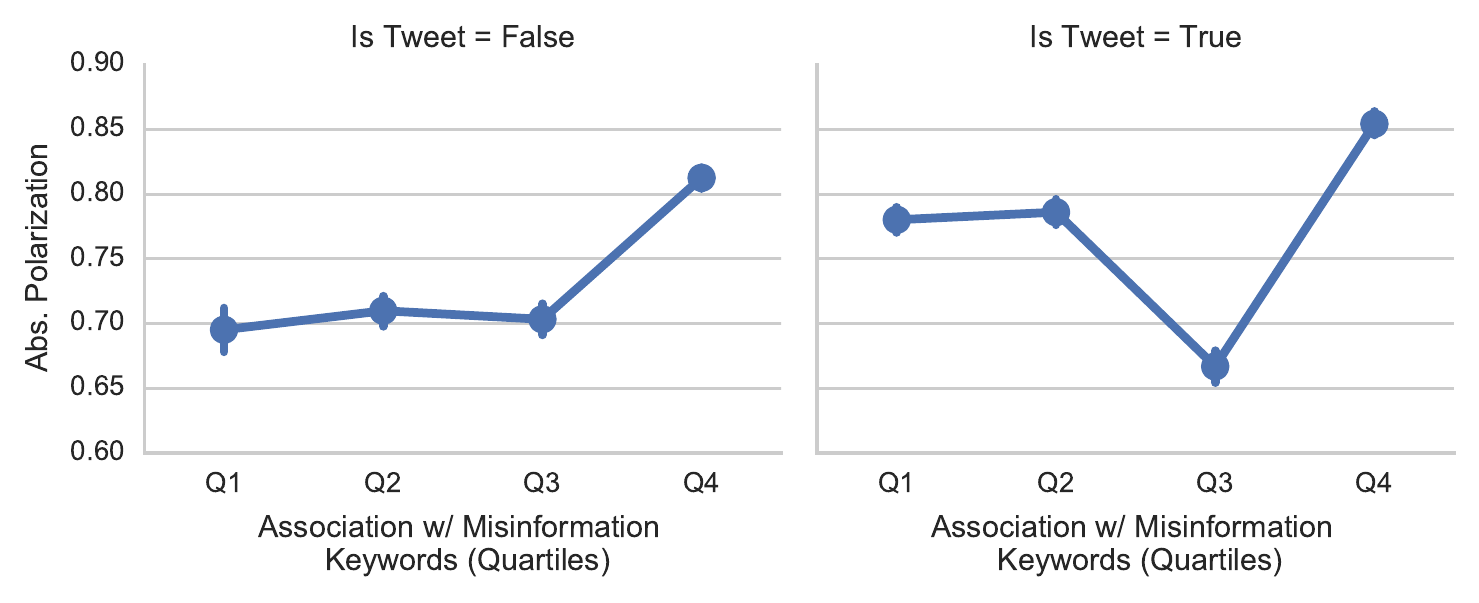}
    \caption{Average polarization per ratio of tweets with the URL containing the misinformation related keywords (quartiles).}
    \label{fig:fn_pp}
    \end{minipage}%

\end{figure*}

Another perspective that can be looked at is how the polarization of URLs changes according to characteristics of the reactions associated to it. We analyze two aspects, namely: what is the impact of the number of reactions surrounding an URL to its polarization, and what is the impact of the percentage of reactions using keywords and hashtags related to fake-news to the polarization of the URL. Ordering the URLs according to these metrics, we plot the average polarization of each one of its quartiles in Figures~\ref{fig:rc_pp} and~\ref{fig:fn_pp}, respectively. We analyze  other tweets and external URLs separately.

Figure~\ref{fig:rc_pp} shows that, in the collected dataset, the increase in the number of reactions has a negative impact in the average polarization of an URL. A better interpretation of these results would require a mapping of the interactions of the users. Figure~\ref{fig:fn_pp} shows an increase of the polarization when URLs are constantly associated with "fake news" related keywords. This contributes to the hypothesis that the "fake new" thematic is a polarizing one.

\begin{figure*}[ht]
    \centering
    \begin{minipage}{.47\textwidth}
        \centering
 \includegraphics[width=\linewidth]{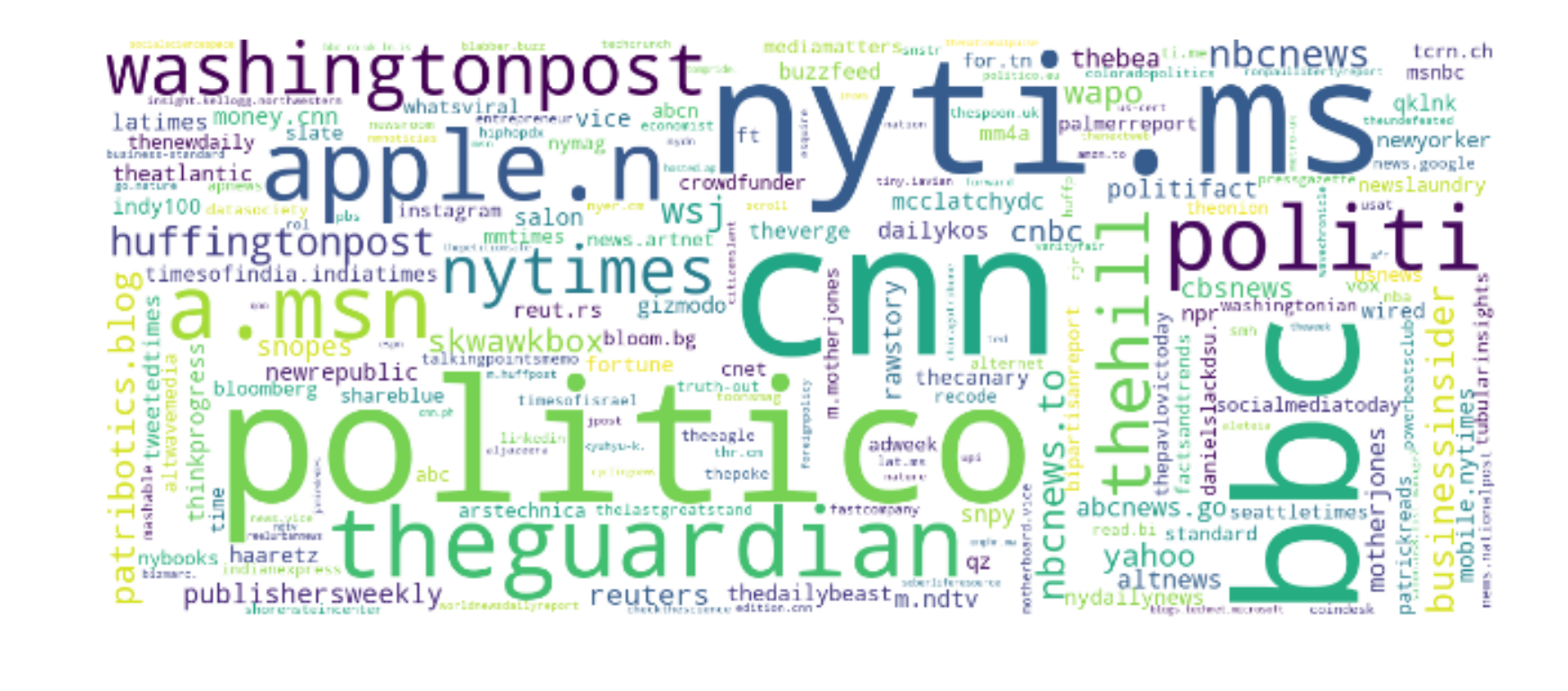}
  \subcaption{Democrat-leaning users.}
    \end{minipage}%
      \hfill
    \begin{minipage}{.47\textwidth}
        \centering
 \includegraphics[width=\linewidth]{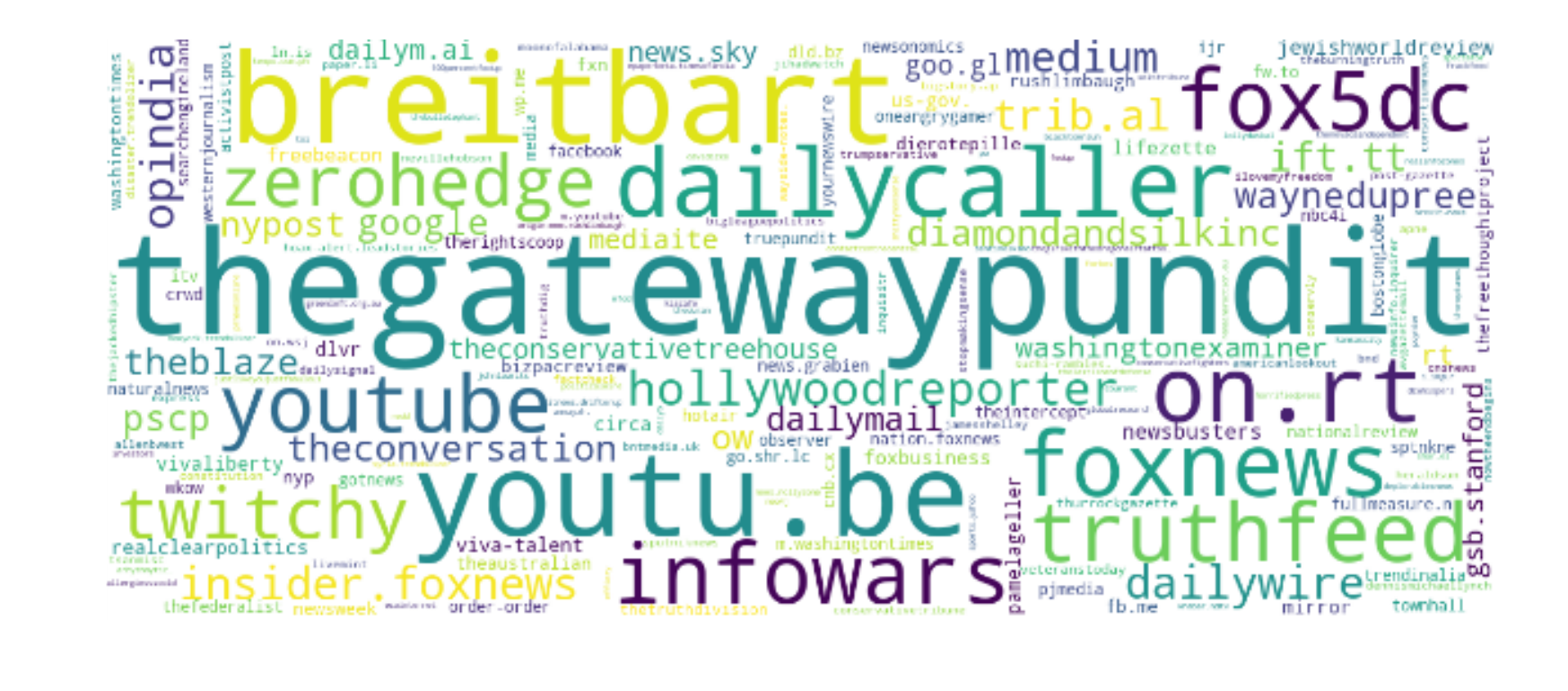}
  \subcaption{Republican-leaning users.}
    \end{minipage}
     \caption{Domain clouds of \#FakeNews-related tweets. Notice the presence of websites with the same ideology as the users in the polarized groups. This indicates that users are reacting to sources they agree with on fake-news related narratives.}
    \label{fig:word_clouds}
\end{figure*}

\subsection{Polarization and URL Domains}
\label{sec:pol_dom}
We generate wordclouds as described in Section~\ref{sec:word_clous_mpac}, and the results can be seen in Figure~\ref{fig:word_clouds}\textit{(a)} for democrat-leaning users and in Figure~\ref{fig:word_clouds}\textit{(b)} for republican-leaning users. Analyzing the typical levels of trust towards different media sources gathered by the Pew Research Center~\cite{mitchell2014political}, we can see that democrat-leaning wordcloud contains domains to news sources such as \textit{The Washington Post} and \textit{The New York Times}, which are reportedly trusted by liberals and distrusted by conservatives. Similarly, the republican-leaning wordcloud contains domains to news sources such as \textit{Breitbart} and \textit{Fox News}, trusted by conservatives and distrusted by liberals.

This implies that polarized group don't directly mention some report or news-piece as fake, but react to links of sources that they agree with on the "fake news" theme. It also indicates that sources that users of a certain political ideology trust have a significant impact on their view of what is fake, as they are citing them as a source rather than the piece of information they believe is misinformation.

\subsection{Analyzing Top Reacted URLs}
Analyzing the tweets which received the most reactions and that co-occurred with "fake news" related keywords allows us to better understand how these are being used. We perform our qualitative analysis by giving and discussing examples of the different stratum we defined and inspected.

Among the randomly selected URLs, for example, the top reacted external URL in the highly polarized stratum $A_k \in [0.67, 1]$ is a news-piece on Michael Flynn being cleared by the FBI as innocent from his relationship with Russian~\cite{nypost}:

\begin{quote}
\textit{New York Post:} FBI clears Michael Flynn in probe linking him to Russia
\end{quote}

\noindent
It is important to notice that Flynn's involvement with Donald Trump's campaign makes this piece of information more favorable to republican-leaning users. Confirming the result obtained with the analysis of the wordclouds in Section~\ref{sec:pol_dom}, however, the result is polarized towards individuals supporting the Republican party. This suggests that users are mainly dismissing a narrative of other media sources that suggested the link of Flynn with  Russia. The terms associated with "fake news" are thus not being employed to denote that a content itself is fake, but denote other pieces of information as false.

Another usage of the term we can find analyzing the more reacted URLs is to refer to news which can be seen as ridicule. One of the most reacted URLs in the less polarized stratum $A_k \in [0, 0.32]$, is about a prisoner who attempted to escape jail  dressed as a woman in Honduras~\cite{telegraph}:

\begin{quote}
\textit{Telegraph:} Prisoner dressed as woman in failed escape bid
\end{quote}

\noindent
This usage, although not necessarily harmful for the political debate, may present a challenge for automated techniques to detect misinformation, if they employ what users in a network such as \textit{Twitter} tag as fake as a feature.

Finally, we can also find instances of users actually tagging actual facts or stories as fake. An example of such case is a highly polarized democrat-leaning news-piece $A_k \in [0.67, 1]$ pointing the rebuttal of a supposedly "fake news" story on the murder of DNC staff:

\begin{quote}
\textit{Raw Story:} Family blasts right-wing media for spreading fake news story about slain DNC staffer as Russia scandal deepens
\end{quote}

We didn't find examples of external URLs of a source known to be trusted by an ideological group being polarized by the opposite political in the our stratified sampling. However, there are cases of tweets where this happens. For example, the following tweet by Donald J. Trump~\cite{trump}~\footnote{We here merely discuss the polarization surrounding a tweet of a public person, not infringing the terms of use of the Twitter's Developer Agreement \& Policy.} is polarized towards democrat-leaning users:

\begin{quote}
\textit{@realdonaldtrump:} The Russia-Trump collusion story is a total hoax, when will this taxpayer funded charade end?
\end{quote}

\noindent
In this case democrat-leaning users may have suggested that what Donald Trump is saying is fake. Previous studies have also shown that retweets of well-known personalities most often denote antagonism~\cite{guerra2017antagonism}.

Although this analysis is not significant to understand  the relevance of these different uses of "fake news" related terms, it provides insight on the plethora of scenarios that co-occur with misinformation-related keywords and hashtags.

\section{Conclusion and Future Work}
\label{conclusion}

This work is a first attempt to observe correlations between political
polarization and the spread of misinformation, in particular "fake news". 
To tackle the practical challenge of having access to a pre-classified set
of fake news articles or tweets, we monitored the external URLs and tweets associated with "fake news" related hashtags and keywords. We searched for tweets reacting to these URLs and  calculated the polarization of the users who reacted to them using an auxiliary more general dataset on politics. We examined the association between polarization and fake news by analyzing the impact of various factors in "fake news" related URLs and users we knew the polarization of. We also analyzed the different sources that are mentioned as "fake" by users and qualitatively described different scenarios where the terminology is applied.

We found that fake-news related debate on Twitter is highly polarized in terms
of the degree of bias of the users that react on "fake news" related URLs and in terms
the different sets of URL domains that democrats and republicans engage with. We also found that, in our dataset, the average polarization was higher when many individuals were tagging a URL as fake. These findings suggest that there is an increase in polarization in the context of content that is related to or perceived as fake news. This tracing of a relationship between polarization and content related to fake news addresses our first research question.

The analysis of the wordclouds of the democrat and republican-leaning users as well as the qualitative examination of the contexts where misinformation-related keywords and hashtags are employed suggest that there is a significant use of "fake news" related keywords to express disagreement. These findings address our second research question. However, the measure to which this happens needs to be assessed quantitatively, as our analyses don't allow the measurement of the impact of such usage.

The impact of polarization in the combat of fake news presents new challenges and opportunities. On one side, if a significant amount of the messages relating content to fake-news expresses disagreement, we have that machine learning methods that use what users indicate as fake as a feature may become biased towards one side of the polarization spectrum. On the other hand, we may use community detection techniques to add polarization as a feature that distinguishes fake and biased, or even find users who are not not extremely polarized, which would have a more trustworthy judgment of what is misinformation.

As future work we want to explore methods to identify different narratives around stories that emerge in distinct polarized communities. This could align  potential "fake news" or extremely polarized articles with others that disprove or reject them, providing a mechanism to regulate the spread and misinformation the polarization it causes in society~\cite{zollo2015emotional,del2016spreading}. Another interesting direction would be to further explore the association of fake news and polarization, finding statements that are proven false by fact-checkers and modeling the complex interactions (such as quotes and replies) between users in social networks such as \textit{Twitter}.

\section*{Acknowledgements}
This work is partially supported by CNPq, CAPES, FAPEMIG, InWeb, MASWEB, and INCT-Cyber.

\bibliographystyle{ACM-Reference-Format}
\bibliography{sigproc} 

\end{document}